\documentstyle[twocolumn,prb,aps,epsfig,graphicx  ]{revtex}

\begin{document}

\title{Fermionic Heisenberg Glasses with BCS Pairing Interaction}

\author{S.\ G.\ Magalh\~{a}es and A.\ A.\ Schmidt}

\address{Departamento de Matem\'{a}tica, UFSM, 97105-900 Santa Maria, RS, Brazil}

\maketitle

\begin{abstract}
We have analyzed a fermionic infinite-ranged quantum Heisenberg spin glass
with BCS coupling in real space in the presence of a magnetic field. It has
been possible to locate the transition line between the normal paramagnetic
phase (NP) and the phase where there is a long range order corresponding to
pair formation in sites (PAIR). The nature of the transition line is also 
investigated. This transition ends at $T_{f}$, the transition temperature 
between NP and spin glass phase (SG) where the static approximation and
replica symmetry ansatz are reliable.

~


\end{abstract}

~

The competition between the superconducting and spin glass ordering has been 
reported in cuprete superconductor [1], heavy fermions [2] and conventional 
superconductors doped with magnetic impurities [3]. Theoretical studies [4] 
in these later systems have been mainly interested in the calculation of the 
superconducting density of states by using a model composed by a conventional 
BCS interaction for the conducting electrons which interact with the localized 
magnetic impurities by a s--d exchange interaction. 

In the present work the interplay between the mechanism responsible by the 
spin glass ordering and the BCS pairing among localized fermions of opposite 
spins has been investigated for the half filling case at mean field level. 
This has been done by using a Hamiltonian with a fermionic Heisenberg spin 
glass and a BCS pairing interaction in real space [5]. This model has been 
obtained from Ref. [4] tracing out by perturbation the conducting fermionic 
degrees of freedom.  

We consider the Hamiltonian given by Eq.\ (A20) of Ref. [5] in the 
presence of a magnetic field $H_{z}$. The problem is formulated in the path 
integral formalism where the spins are 
represented by bilinear combinations of Grassmann fields and random
coupling among the spins are infinite ranged with a Gaussian distribution 
with zero mean and variance given by Ref. [6]. The 
disorder is treated in the context of the replica method with an important 
difference compared with its classical counterparts: in the quantum case the 
diagonal component of the spin glass order parameter is no longer constrained 
to unity. In fact, the location of other transitions depends on this component 
even when the non-diagonal component is zero. The magnetic field in the $z$ 
direction separates the order parameters in two groups: transversal to the 
field and parallel to it [7]. The transversal 
non-diagonal spin glass order parameter is taken as null. The static 
approximations and the replica symmetry ansatz is used for the remaining 
order parameters. This procedure is reliable up to $T_{f}$ (the transition 
temperature between the spin glass (SG) and the normal-paramagnetic (NP) 
phases) which moves down as the field strength increases. Therefore, the 
Grand Canonical potential (details will be given elsewhere [8]) 
can be found for the half-filling case as
\begin{eqnarray}
\Omega=2\beta J^2(R^2+R_z-Q_z)
+\frac{g}{4}|\eta|^{2} \nonumber\\ 
-\frac 1\beta \int_{-\infty }^{+\infty }Dw\ln(I_{\beta}) 
\label{2.25}
\end{eqnarray}

\begin{eqnarray}
I_{\beta}=\int_0^{+\infty}uDu\int_{-\infty}^{+\infty}Dv
[\cosh(\frac{\beta g}{2}|\eta|) \nonumber\\
+\cosh(\beta|\vec{h}|)] 
\label{2.26}
\end{eqnarray}
where $|\vec{h}|=J\sqrt{2Ru^{2}+\theta^{2}}$ and 
$\theta=[v\sqrt{2(R_{z}-Q_{z})}$ 
$+w\sqrt{2Q_{z}}+H_{z}/(2J)]$. In both equations $Du=du\exp(-u^{2}/2)/\sqrt{2\pi}$
and $\beta=1/T$ and $J$ is given by Eq.\ (3) of Ref.\ [5].

In Eq.\ (\ref{2.25}), $|\eta|$ is the PAIR order parameter, $Q_{z}$ and 
$R_{z}$ are, respectively, the non-diagonal and the diagonal component 
of the spin glass order parameter parallel to the field $H_{z}$ and $R$ 
is the diagonal component transversal to it. The saddle point equations 
for $R$, $Q_{z}$, $R_{z}$ and $|\eta|$ follow from Eq.\ (\ref{2.25}).

A phase diagram can be obtained in $T$--$g$ ($g$ is the pairing strength) 
space for different values of $H_{z}$ by solving the resulting equations
for $R$, $Q_{z}$, $R_{z}$ and $|\eta|$ (see Fig.\ (1)). 

\vfill

\begin{figure}
\centering
\vspace*{1mm}
\includegraphics[scale=0.55,viewport=0 0 420 290]{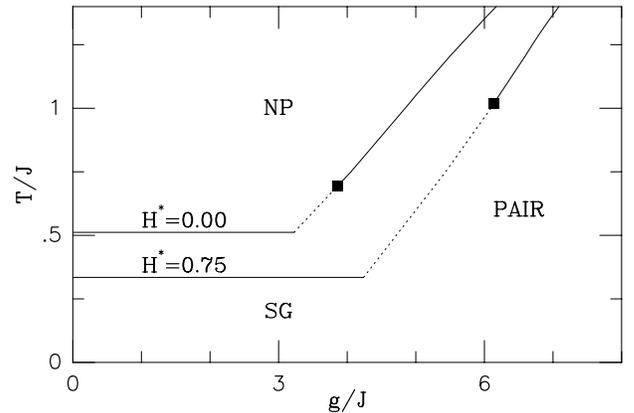}
\vspace*{1mm}
\caption{Phase diagram as a function of temperature and pairing 
coupling $g/J$ for two values of $H^{*}$ where  $H^{*}=H_{z}/J/(8)^{1/2}$. 
Solid lines indicate second order transitions while dotted lines 
indicate a first order transition. Tricritical points are shown as 
filled squares.}
\end{figure} 

\newpage

If $H_{z}=0$, for high $T$ and small $g$ there is no long range order 
($|\eta|=0$) that corresponds to NP phase. For high $T$ and $g$, 
$|\eta|\neq 0$ which corresponds to the PAIR phase. For the transition 
line, it has been found a crossover from a continuous behavior to a  
sharp one at $(T_{tc},g_{tc})$. The changing in the behavior of 
$|\eta|$ is shown in Fig.\ (2). 

The parameter $Q_{z}$ is null and $R=R_{z}$ shows a 
crossover from continuous to discontinuous behavior as $|\eta|$ does. 
If $H_{z}\neq 0$, the PAIR phase only exists for larger values of $g$. 
The point ($T_{tc}$, $g_{tc}$) is moved up and the region where the 
transition is the first order type increases. The nature of ($T_{tc}$, 
$g_{tc}$) as a tricritical point has been confirmed by following the 
same procedure of Ref. [5]. The behavior of $R$, $R_{z}$ and $Q_{z}$ 
changes strongly as shown in the Fig.\ (3). 

The temperature $T_{f}$ can be found from the expansion of the Grand Canonical 
potential up to the quadratic term of the non-diagonal component transversal 
to the field of the spin glass order parameter [7]. The condition that the 
coefficient of the quadratic term vanishes gives $T_{f}/J=4R$ (in the static 
approximation) which must be solved with the equations for the order parameters.

To conclude, this work has analyzed a fermionic Heisenberg spin glass model 
with a BCS pairing among local fermions which has been solved by reduction to 
a one site problem. A transition line separating the NP and PAIR phases has been
obtained having a tricritical point depending on $H_{z}$ from where second
order transition occurs for higher values of $g$ and first order transition
occurs for lower values of $g$.

\begin{figure}
\centering
\vspace*{10mm}
\includegraphics[scale=0.55,viewport=0 0 420 310]{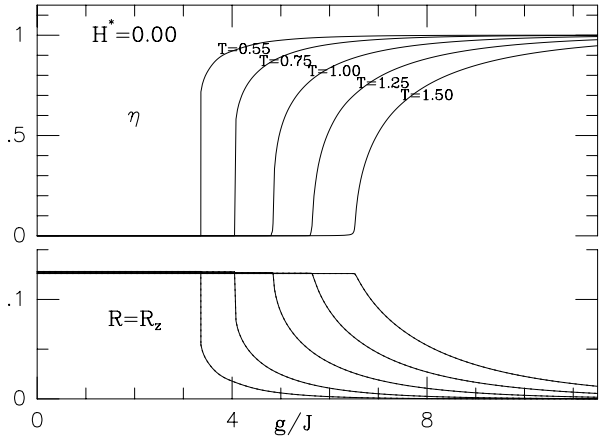}
\vspace*{1mm}
\caption{Dependence of the order parameters $\eta$, $R$, $R_z$ 
and $Q_z$ for $H^{*}=0.00$ as a function of $g/J$ for several values of 
temperature. For $H^{*}=0.00$, $Q_z$ is always null.}
\end{figure}

\newpage

\begin{figure}
\centering
\vspace*{0mm}
\includegraphics[scale=0.55,viewport=0 0 420 500]{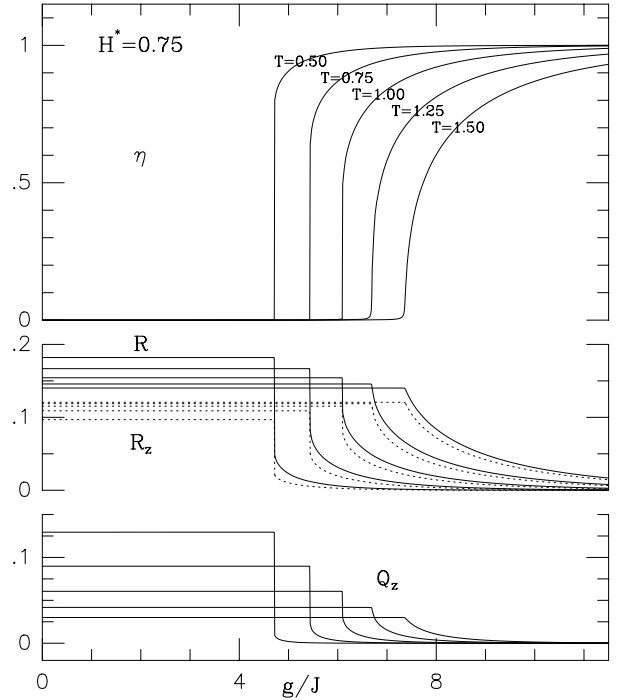}
\vspace*{1mm}
\caption{Same as Fig.\ 2 for $H^{*}/J=0.75$.}
\end{figure} 

\vspace*{14mm}

\noindent{\bf References}

~

\noindent[1] H.\ Spille et al., J.\ Magn.\ Magn., {\bf 76/77 }, 539(1988).

\noindent[2] F.\ C.\ Chou et al., Phys.\ Rev.\ Lett., {\bf 75}, 2204(1995). 

\noindent[3] D.\ Davidov et al., J.\ Phys.\ F.: Metal.\ Phys., {\bf 7} L47 
\phantom{[3] }(1977).

\noindent[4] M.\ J.\ Nass, K.\ Levin and G.\ S.\ Grest, Phys.\ Rev.\ B, 
\phantom{[4] }{\bf 23} 1111 (1981).

\noindent[5] S.\ G.\ Magalh\~aes and Alba Theumann, Europ.\ Phys.\ 
\phantom{[5] }J.\ B, {\bf 9}, 5 (1999).

\noindent[6] Alba Theumann, Phys.\ Rev.\ B, {\bf 33}, 559 (1986).  

\noindent[7] Yadin Y.\ Goldshmidt and Pik-Yin Lai, Phys.\ Rev.\ B, 
\phantom{[7] }{\bf 43}, 11434 (1991).

\noindent[8] S.\ G.\ Magalh\~{a}es and A.\ A.\ Schmidt, Phys.\ Rev.\ B, 
\phantom{[7] }(accepted for publication).

\end{document}